\documentclass[useAMS,usenatbib,usegraphicx]{mn2e} 
\usepackage{amssymb} 
\usepackage{times}
\usepackage{aas_macros}
\usepackage[usenames]{color}

\title[Amplitude Modulation in KIC\,7106205]{Pulsational Frequency and Amplitude Modulation in the $\delta$\,Sct star KIC\,7106205} 

\author[D. M. Bowman and D. W. Kurtz] {Dominic M. Bowman\thanks{Email: dmbowman@uclan.ac.uk} and Donald W. Kurtz \\
Jeremiah Horrocks Institute, University of Central Lancashire, Preston PR1 2HE, UK\\ }

\date{\today}

\begin{document} 

\maketitle 

\begin{abstract} 
Analysis of the \textit{Kepler} $\delta$\,Sct star KIC\,7106205 showed amplitude modulation in a single pressure mode, whilst all other pressure and gravity modes remained stable in amplitude and phase over the 1470\,d length of the dataset. The \textit{Kepler} dataset was divided into a series with time bins of equal length for which consecutive Fourier transforms were calculated. An optimum fixed frequency, calculated from a least-squares fit of all data, allowed amplitude and phase of each pulsation mode for each time bin to be tracked. The single pressure mode at $\nu$\,=\,13.3942\,d$^{-1}$ changed significantly in amplitude, from 5.16\,$\pm$\,0.03 to 0.53\,$\pm$\,0.06\,mmag, but also varied quasi-sinusoidally in phase, with a characteristic period similar to the length of the dataset. All other p and g\,modes were stable in both amplitude and phase, which is clear evidence that the visible pulsation mode energy is not conserved within this star. Possible causes of the observed amplitude and phase modulation and the missing mode energy are discussed.
\end{abstract} 

\begin{keywords}
asteroseismology -- stars: oscillations -- stars: variables: $\delta$\,Sct -- stars: individual (KIC\,7106205)
\end{keywords} 


\section{Introduction} 
\label{section: Introduction}

The recent revolution in space photometry, facilitated by the MOST, CoRoT and \textit{Kepler} space missions, has dramatically improved our understanding for a variety of different types of stars. The vast increase in high quality data has driven remarkable advances into understanding both convection and the mechanisms that drive stellar pulsations in A and F-type variable stars \citep{Guzik2000a, Dupret2004, Dupret2005}, i.e., the $\delta$\,Sct and $\gamma$\,Dor stars.

Asteroseismology is a rapidly expanding field of astrophysics, in which stellar pulsations and the internal structures of stars are studied from periodic changes in brightness and radial velocity of their surfaces. With more than 4 yr of high duty cycle data from the \textit{Kepler} space telescope, we are able to probe frequency and amplitude modulation of pulsation modes in $\delta$\,Sct and $\gamma$\,Dor stars. The high-precision values of frequency, amplitude and phase for individual pulsation modes in \textit{Kepler} stars yield cleaner and richer amplitude spectra than can be obtained from a ground-based telescope and, most importantly, a better insight into the various pulsation excitation mechanisms.

A little more than a decade ago, \citet{Rod2001} commented in their catalogue of $\delta$\,Sct and related stars, that ``\textit{severe selection effects exist}'' in ground-based observations, which made any analysis requiring high-precision values, such as mode identification, difficult. However, \textit{Kepler} data are not limited by many of the observational biases encountered when using ground-based surveys, such as poor signal-to-noise and aliasing caused by large gaps in the data. In this paper, we demonstrate the superiority of space photometry, and in particular \textit{Kepler} data, for asteroseismology and the study of pulsational amplitude modulation. 

	\subsection{The \textit{Kepler} Mission}
	\label{subsection: Kepler Mission}

	The \textit{Kepler} space telescope, launched in 2009 March, has recorded photometric observations of more than 190\,000 stars at the $\umu$mag precision level \citep{Koch2010}. The \textit{Kepler} telescope is in a 372.5-d Earth-trailing orbit and the field of view covered approximately 115\,deg$^{2}$ in the constellations of Cygnus and Lyra. The primary goal of the \textit{Kepler} mission was to locate Earth-like planets in the habitable zone of their host star using the transit method \citep{Borucki2010}. 
	
	The main mission came to an end in 2013 May when the telescope lost the function of a second reaction wheel that was vital for the still-pointing of the spacecraft. Since the failure, the mission parameters have been changed and the telescope is now observing in the plane of the ecliptic, because in this position solar torques can be balanced so that spacecraft drift can be controlled with minimum loss of propellent \citep{Haas2014}. The new mission has been named `K2', and provides a variety of targets, such as young stars and star-forming regions, supernovae, white dwarfs, and the Galactic centre \citep{Haas2014}. The telescope may no longer be collecting data on the original target list, but there is more than 4\,yr of high quality data available to study on a variety of different stellar objects -- a goldmine for scientific discoveries.

	\textit{Kepler} data are available in two formats, long and short cadence (hereafter called LC and SC respectively). \textit{Kepler's} primary mission goal required a short integration time in order to maximise the number of data points for each transit of a planet in front of its host star. This was achieved by the SC data, which had an integration time of 58.5-s. However, this limited the number of stars that were observed at this cadence to 512 at any one time \citep{Gilliland2010}. LC data had an integration time of 29.5-min and consequently, the maximum number of LC targets was 170\,000 \citep{Jenkins2010b}. Therefore, a compromise between the number and temporal resolution of \textit{Kepler} targets was chosen for a given observing time. Most target stars were chosen to fulfil \textit{Kepler's} primary goal of detecting planetary transits, but approximately 1 per cent of all targets were reserved solely for asteroseismology \citep{Gilliland2010}.
	
	LC data are grouped in quarters, denoted by Q[n], where n\,=\,0\,--\,17, which lasted for approximately 93\,d, although Q0, Q1 and Q17 were approximately 10, 30 and 30\,d, respectively. SC data are denoted by Q[n]M[m], where m\,=\,1,\,2 or 3. Every quarter, \textit{Kepler} had to roll 90 degrees in order to keep its solar panels pointing towards the Sun and the radiator for keeping it cool pointing into deep space. Therefore, a star rotated its position on the focal plane 4 times throughout an entire \textit{Kepler} year. For a detailed review concerning the differences in SC and LC asteroseismology using \textit{Kepler} data and also the differences in the data processing pipelines, see \citet{Murphy2012a}. 
	
	
\begin{table*}
	\centering
	\caption[]{Stellar parameters listed for KIC\,7106205 in the \textit{Kepler} Input Catalogue \citep{Brown2011} and the revised values given in \citet{Huber2014}.}
		\begin{tabular}{c c c c c c c}
		\hline
		\multicolumn{1}{c}{} & \multicolumn{1}{c}{T$_{\mathrm{eff}}$} & \multicolumn{1}{c}{$\log g$ } & \multicolumn{1}{c}{[Fe/H]} & \multicolumn{1}{c}{Radius} & \multicolumn{1}{c}{m$_{\rmn{v}}$} & \multicolumn{1}{c}{Contamination} \\
		\multicolumn{1}{c}{} & \multicolumn{1}{c}{(K)} & \multicolumn{1}{c}{(cm\,s$^{-2}$)} & \multicolumn{1}{c}{(dex)} & \multicolumn{1}{c}{(R$_{\sun}$)} & \multicolumn{1}{c}{(mag)} & \multicolumn{1}{c}{($\%$)} \\
		\hline
		KIC \citep{Brown2011} & 6960 $\pm$ 150 & 4.05 $\pm$ 0.15 & -0.01 $\pm$ 0.15 & 1.78 $\pm$ 0.89 & 11.46 & 0.005 \\ 
		\\
		Revised KIC \citep{Huber2014} & 6900 $\pm$ 140 & 3.70 $\pm$ 0.15 & 0.32 $\pm$ 0.15 & 3.23 $\pm$ 0.61 & - & -  \\
		\hline
		\end{tabular}
	\label{table: stellar parameters}
\end{table*}
 

	\subsection{Delta Scuti Stars}
	\label{subsection: Delta Scuti Stars}

	Delta Scuti stars are the most common group of pulsating A-type stars, which lie at the intersection of the main sequence and the classical instability strip on the Hertzsprung-Russell (HR) diagram. Whilst on the main sequence, they range from A2 to F0 in spectral type and lie between 7000 $\leq T_{\rmn{eff}} \leq$ 9300 K \citep{Uytterhoeven2011}. 
	
	The $\delta$\,Sct pulsations are excited by the $\kappa$-mechanism, which operates in the He\,\textsc{ii} ionisation zone, and consequently pressure (p) modes and gravity (g) modes are observed \citep{Chevalier1971a}. The changes in opacity, and therefore pressure, of a parcel of gas relative to its surroundings are analogous to a piston in a heat engine cycle. The result is a periodic expansion and reduction in the radius of the parcel of gas giving rise to the nomenclature of pressure modes. Typical periods for p\,mode pulsations in $\delta$\,Sct stars range from 15\,min to 5\,h \citep{Uytterhoeven2011}. See \citet{Breger2000a} for a thorough review of $\delta$\,Sct stars.

	\subsection{Gamma Doradus Stars}
	\label{subsection: Gamma Doradus Stars}
	
	Gamma Doradus stars have only been recognised as a distinct category of pulsating objects in the last two decades \citep{Kaye2000}. They have similar effective temperatures and luminosities to $\delta$\,Sct stars, such that both types of star occupy overlapping regions on the HR\,diagram \citep{Uytterhoeven2011}. Most $\gamma$\,Dor stars observed by \textit{Kepler} lie on the main sequence and range between F5 to A7 in spectral type and between 6500 $\leq T_{\rmn{eff}} \leq$ 7800 K \citep{Uytterhoeven2011}. However, using \textit{Hipparcos} photometry, \citet{Handler1999} found that $\gamma$\,Dor stars lie between 7200 $\leq T_{\rmn{eff}} \leq$ 7700 K, which illustrates that a difference in the definition of the red edge of the instability strip can cause confusion between distinguishing $\gamma$\,Dor and $\delta$\,Sct stars. Therefore, knowledge of the star's pulsation frequencies is essential in order to classify the pulsator type.
		
	Even though they have similar spectral parameters to $\delta$\,Sct stars, pulsations observed in $\gamma$\,Dor stars are driven by a different mechanism: the flux modulation or flux blocking mechanism \citep{Guzik2000a, Dupret2004, Griga2005}. Pulsation driving requires the local convective timescale at the base of the convection zone to be similar to, or longer than the pulsation period, such that convection cannot adapt quickly enough to damp the pulsations \citep{Guzik2000a}. These buoyancy driven, g\,mode pulsations have much longer periods than p\,modes observed in $\delta$\,Sct stars and range between 8\,h and 3\,d \citep{Uytterhoeven2011}.
	
	\subsection{Hybrid Stars}
	\label{subsection: Hybrid Stars}
	
	Due to the unprecedented photometric precision, high duty-cycle and length of \textit{Kepler} data, more pulsation modes have been observed in a variety of stellar types than was previously possible. Some stars have been found to exhibit modes in both the $\gamma$\,Dor regime $(\nu<4\,$d$^{-1})$ and the $\delta$\,Sct regime $(\nu\geq5\,$d$^{-1})$. Attempts have been made to catalogue the many different types of behaviour that these stars present. For example, it has been suggested that $\delta$\,Sct$/$$\gamma$\,Dor hybrids and $\gamma$\,Dor$/$$\delta$\,Sct hybrids are two distinct groups \citep{Griga2010}.
	
	Classifying a hybrid star often requires a visual inspection of the light curves for the star, which can be a lengthy process. Nonetheless, it is clear that a significant fraction of A- and F-type stars exhibit at least some form of hybrid behaviour. The once separate groups of $\gamma$\,Dor and $\delta$\,Sct stars overlap more than was previously thought as some stars show pulsations excited by the two different mechanisms \citep{Uytterhoeven2011}. From analysis of the complete \textit{Kepler} dataset catalogue, at least 23 per cent of all A- and F-type stars can be classed as hybrid stars \citep{Uytterhoeven2011}. 
	
	These stars are ideal test candidates to study pulsation energy conservation between the p and g\,modes, and also between the modes and the driving zones, for both opacity and convective driving. If, for example, the visible pulsation energy is conserved within a star, then any decrease in mode amplitude in one or more modes would result in any number of other modes increasing in amplitude and vice versa. This concept has not been thoroughly investigated before now, and it is important to understand whether a star's pulsation energy budget is constant, especially on such a short time scale as 4 yr. If a star does not conserve its pulsation energy, then it may be transferred to either the convection or ionisation driving zones, or to another damping region.
 
 
\section{Amplitude Modulation}
\label{section: Amplitude Modulation}
	
	The vast number of stars available to study within the \textit{Kepler} dataset has allowed the variety of observed stellar pulsations to be probed. Prior to space-based missions that were useful to asteroseismology, ground-based observations of variable stars had been limited to observing pulsations with amplitudes larger than a few mmag. Even so, some $\delta$\,Sct stars contained some constant and some extremely variable pulsation modes \citep{Breger2009}. There are few examples of stars that demonstrate this behaviour, and some case studies are given below.
	
	From theoretical studies of $\delta$\,Sct stars, mode-coupling is expected between different combinations of frequencies \citep{Dziembowski1982}. Parametric resonance instability can occur in which the instability of a linearly driven mode at $\nu_1$ causes the growth of two modes at $\nu_2$ and $\nu_3$ such that $\nu_1 \approx \nu_2 + \nu_3$, satisfying the resonance condition \citep{Dziembowski1982}. The decay of either a linearly driven p or g\,mode into 2 g\,modes is most likely as linearly driven modes usually have low radial orders \citep{Dziembowski1982}. Observationally, this is difficult to confirm due to the extremely low g\,mode amplitudes and the complexities of different mode-coupling mechanisms at work within a star. However, with the high precision of the \textit{Kepler} data, we can now begin to address this question.
	
	\subsection{Case Studies}
	
	In a recent paper, \citet{Breger2014} analysed a rapidly rotating A-type star, KIC\,8054146, and observed mode-coupling of two `parent' modes that produced a single `child' mode, as predicted by the resonance condition given by \citet{Dziembowski1982}. The amplitude changes of a combination mode are predicted to follow the product of the amplitude changes in the two parent modes, which was subsequently used to identify which modes were the parent and child modes \citep{Breger2014}. KIC\,8054146 contained 349 statistically significant frequencies in the range $0<\nu\leq200$\,d$^{-1}$, including three separate `families' of frequencies whose amplitude variations of the low frequency members correlated with the high frequency ones \citep{Breger2014}. Thus, energy seemed to be conserved between the visible pulsation modes in KIC\,8054146.
	
	Analysis of the $\delta$\,Sct star 4\,CVn over 40 years showed modes that appeared and disappeared unpredictably, which illustrated mode stability in A-type stars \citep{Breger2000b, Breger2009}. A single p\,mode at $\nu$\,=\,7.375\,d$^{-1}$ decreased in amplitude from 15\,mmag in 1974, to 4\,mmag in 1976 and to 1\,mmag in 1977, after which a phase jump occurred and the mode began increasing in amplitude again \citep{Breger2000b}. Observations suggested that other frequencies were strongly coupled to the $\nu$\,=\,7.375\,d$^{-1}$ mode and \citet{Breger2000b} speculated that power was being transferred between modes by mode-coupling and suggested a few plausible  possibilities:
	
	\begin{enumerate}
	
	\item \textit{Mode Beating:} a simple beating model is supported by the observed phase change, which suggested that two modes of similar frequency and amplitude were beating against each other. However, the poor quality of the fit from this model led to this hypothesis being rejected \citep{Breger2000b}.
	
	\item \textit{Re-excitation:} the mode decayed and soon afterwards was re-excited with a completely random phase. However, this does not require the phase change of half a cycle that was observed \citep{Breger2000b}.
	
	\item \textit{Stellar Cycle:} the apparent periodic behaviour of the mode amplitude could be evidence of a stellar cycle as cyclic periodicity in mode amplitude cannot be explained by stellar evolution \citep{Breger2000b}. 
	
	\end{enumerate}
	
	With more data, \citet{Breger2009} found that the mode variability could be fitted with a period of decades - far longer than the current dataset available for the star. Therefore, without seeing even a single cycle for this mode, it was difficult to test the stellar cycle hypothesis. Therefore, the amplitude modulation in 4\,CVn remains an unsolved problem.

A different form of amplitude modulation was seen in the Am star KIC\,3429637 (HD\,178875), which showed continual growth in mode amplitude across 2\,yr of the \textit{Kepler} dataset \citep{Murphy2012b}. It was shown that two of the three most prominent modes grew in amplitude, whilst the other decreased -- changes in all three modes required different functional forms. \citet{Murphy2012b} demonstrated that this was not an instrumental effect, as all modes would decrease or increase with the same functional form if the modulation was instrumental. \citet{Murphy2012b} concluded that real-time evolution of the star across 2\,yr of observations was seen. As the star evolved, the changes in the different pulsation cavities modulated the observed pulsation amplitudes \citep{Murphy2012b}.

	Even though they may act as useful case studies, it is difficult to draw conclusions concerning amplitude modulation from a few stars, conclusions that may have been speculative and specific to the case study in question. This is strong motivation for studying this phenomenon in detail using the much larger number of stars showing amplitude modulation within \textit{Kepler} data. Clearly, it is important to understand the various interactions between different pulsation modes, even those that are driven by different excitation mechanisms, which reinforces the importance of studying hybrid stars and understanding the stellar pulsations they produce. To that end, the following hypothesis has been tested: \textit{Is the visible pulsation energy conserved within KIC\,7106205?} From this, an investigation of energy conservation and mode-coupling followed.



	\begin{figure*}
	\centering
		\includegraphics[width=0.49\linewidth,angle=0]{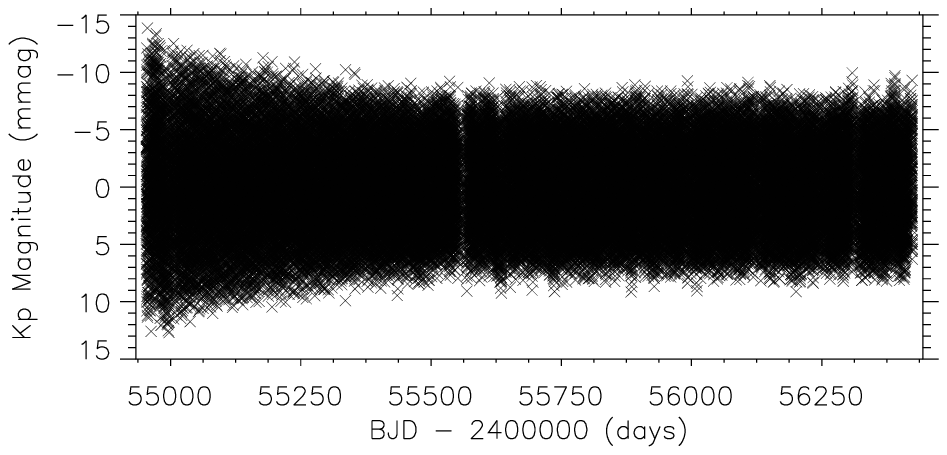}	
		\includegraphics[width=0.49\linewidth,angle=0]{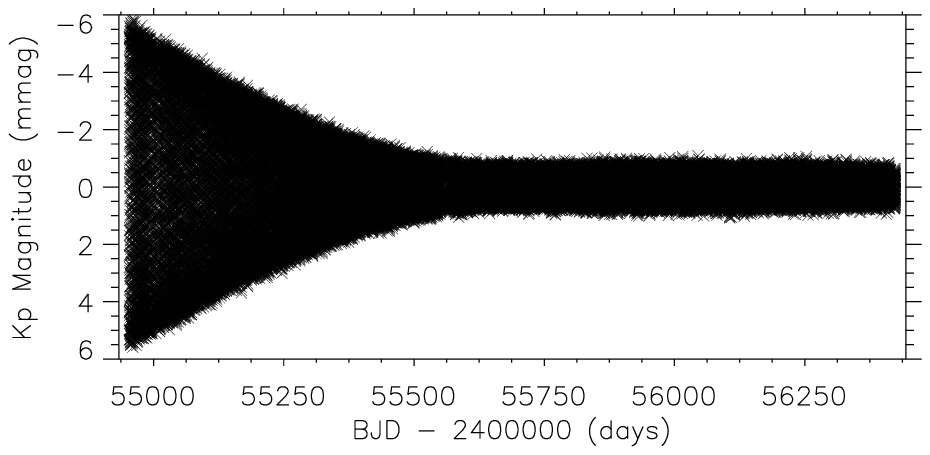}
		\includegraphics[width=0.99\linewidth,angle=0]{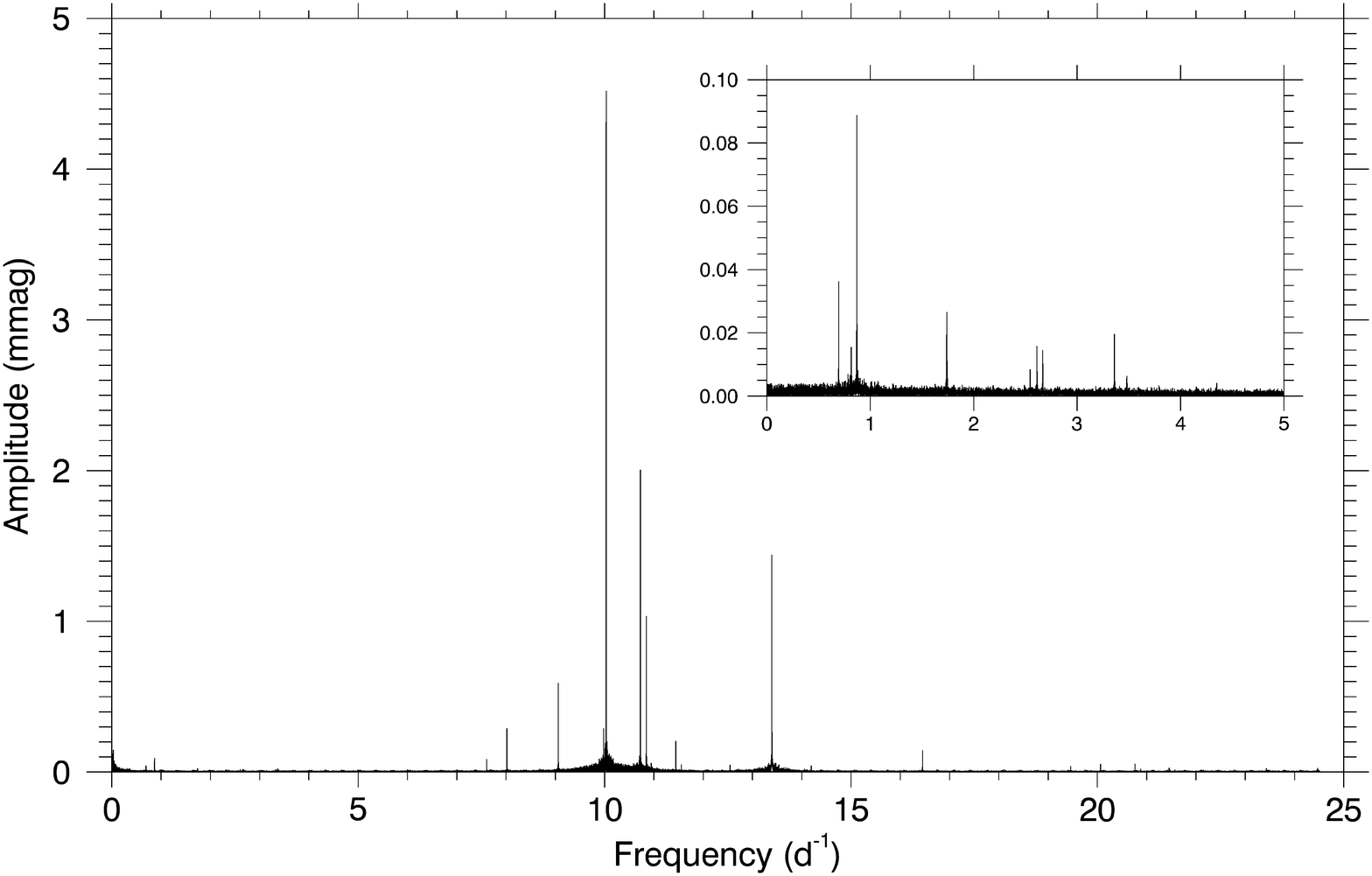}
	\caption{Top Left Panel: The light curve of all LC \textit{Kepler} data for KIC\,7106205. Top Right Panel: The light curve of KIC\,7106205 with all significant peaks prewhitened down to 8\,$\umu$mag except for $\nu=$13.3942\,d$^{-1}$. Bottom Panel: The discrete Fourier transform of all LC \textit{Kepler} data for KIC\,7106205 out to the Nyquist frequency of 24.51\,d$^{-1}$. The sub-plot in the bottom panel shows the g\,mode frequency regime which contains several low amplitude frequencies.}
	\label{figure: all quarters}
	\end{figure*}
	

\section{Data Retrieval: From Star to Screen}
\label{section: Data Retrieval}

In order to maximise the number of case studies of amplitude modulation, and to create a catalogue of modulated $\delta$\,Sct and $\gamma$\,Dor stars, a data retrieval and processing method has been automated for all stars between 6500\,$\leq\,$T$_{\rmn{eff}}\leq$ 20\,000\,K in the \textit{Kepler} Input Catalogue (KIC; \citealt{Brown2011}). Since \textit{Kepler}'s launch, the resultant photometric time-series have been through several versions of the different data processing pipelines, the latest being the multi-scale Maximum A Posteriori (msMAP) implementation of the Pre-search Data Conditioning (PDC) module\footnote{ Kepler Data Notes: https://archive.stsci.edu/kepler/data$\_$release.html }. Data are available from MAST\footnote{MAST website: http://archive.stsci.edu/kepler/} in {\sevensize \sc FITS} file format, which were downloaded for stars between 6500\,$\leq\,$T$_{\rmn{eff}}\leq$ 20\,000\,K in the KIC and stored locally. These {\sevensize \sc FITS} files were then processed to produce time series in reduced Barycentric Julian Date (BJD - 2400000) and normalised magnitudes for each quarter of LC and month of SC data for every star. A pipeline for producing Fourier transforms for each quarter of LC data was also automated and stars that met the criteria of significant amplitude modulation of at least one pulsation mode were flagged for further study.

	The light curve of KIC\,7106205 in the top left panel of Fig.\,\ref{figure: all quarters} shows reduction in the amplitude of variability of the star on a timescale less than the length of the dataset, which spans a total of 1470\,d (4\,yr). No data points have been deleted and no gaps have been interpolated in the time series for this star. 
	
	The KIC parameters for KIC\,7106025 and the revised values from \citet{Huber2014}, with their respective errors, are given in Table\,\ref{table: stellar parameters}. The values of T$_{\rm eff} $ (6900\,K) from both of these catalogues are consistent and characterises KIC\,7106205 as an early F-type star. KIC\,7106205 demonstrates a conclusion stated in \citet{Huber2014} that stars with T$_{\rm eff}\leq7000$ listed in the original KIC have robust values for T$_{\rm eff}$. The revised values of $\log g$ and metallicity are sourced from spectroscopy and demonstrate the inconsistencies between spectroscopy and photometry for deriving $\log g$ and metallicity values.
	
	In their overview of A- and F-type stars, \citet{Uytterhoeven2011} classify KIC\,7106205 as a pure $\delta$\,Sct star (i.e. shows no evidence of hybrid behaviour) and also a possible binary system. However, the analysis done by \citet{Uytterhoeven2011} used only Q0 and Q1 of the \textit{Kepler} dataset and as such, the resultant frequency resolution with only 40\,d of data was 0.025\,d$^{-1} (\approx$\,30\,$\umu$Hz). With 1470\,d of LC data available for KIC\,7106205, consisting of 65\,308 data points, the frequency resolution was 6.8\,$\times 10^{-4}$\,d$^{-1} (\approx$\,8\,nHz), and thus frequencies from stars with 1470\,d of \textit{Kepler} data are well-resolved. As for possible binarity, the target pixel files were also extracted from MAST, and it was clear that there was negligible contamination from background sources. Contrary to \citet{Uytterhoeven2011} we found in this analysis that KIC\,7106205 is a single star. There is no evidence of the pulsational frequency modulation that would be found in a binary system \citep{Shibahashi2012}.
	
	The Nyquist frequency of LC \textit{Kepler} data is $1/(2\Delta t)$\,=\,24.51\,d$^{-1}$, where $\Delta t=29.5$-min is the cadence of the data. On board \textit{Kepler} the data were regularly sampled, but due to the orbit of the spacecraft, barycentric corrections were made to the time series in order to correct for the difference in light arrival time of the photons to the barycentre of the solar system and to the telescope respectively \citep{Murphy2013}. This correction modified the time stamps in the time series to be in BJD and resulted in a cadence which was not constant. An advantage of the unequally-spaced data allows one to easily identify Nyquist aliases using super-Nyquist asteroseismology \citep{Murphy2013}.


	
\begin{table*}
	\centering
	\caption[]{ Frequencies with their associated amplitudes, phases with respective errors, extracted using a least squares fit of 1470\,d of data for KIC\,7106205. Linear combination frequencies are labelled and pulsation constants are given for the identified real frequencies. The table only gives the g and p\,mode frequencies with amplitudes $\geq$ 0.01\,mmag and $\geq$ 0.1\,mmag respectively.}
	\small
		\begin{tabular}{c r r r c c}
		\hline
		\multicolumn{1}{c}{} &\multicolumn{1}{c}{$\nu$} & \multicolumn{1}{c}{Amplitude} & \multicolumn{1}{c}{Phase} & \multicolumn{1}{c}{Comment} & \multicolumn{1}{c}{Pulsation Constant} \\	
		\multicolumn{1}{c}{} &\multicolumn{1}{c}{(d$^{-1}$)} & \multicolumn{1}{c}{(mmag)} & \multicolumn{1}{c}{(rad)} & \multicolumn{1}{c}{}  & \multicolumn{1}{c}{(d)} \\		
		\hline
		g$_1$	&	0.6949  	& 	0.036 $\pm$ 0.006 	& 	0.231 $\pm$ 0.153	&	g$_1$ = g$_8$ - g$_7$	&	Q = 0.3247	\\
		g$_2$	&	0.8153  	& 	0.016 $\pm$ 0.006 	& 	2.993 $\pm$ 0.357	&	g$_2$ = g$_8$ - g$_5$	&	Q = 0.2767	\\
		g$_3$	&	0.8697  	& 	0.089 $\pm$ 0.006 	& 	2.986 $\pm$ 0.063	&						&	Q = 0.2594	\\
		g$_4$	&	1.7397  	& 	0.027 $\pm$ 0.006 	& 	-3.029 $\pm$ 0.209 	&						&	Q = 0.1297 	\\
		g$_5$	&	2.5464  	& 	0.008 $\pm$ 0.006 	& 	2.014 $\pm$ 0.659 	&	g$_5$ = g$_8$ - g$_2$	& 	Q = 0.0886	\\
		g$_6$	&	2.6108  	& 	0.016 $\pm$ 0.006 	& 	-2.280 $\pm$ 0.349 	&						&	Q = 0.0864	\\
		g$_7$	&	2.6669  	& 	0.015 $\pm$ 0.006 	& 	-1.791 $\pm$ 0.381 	&	g$_7$ = g$_8$ - g$_1$	&	Q = 0.0846	\\
		g$_8$	&	3.3618  	& 	0.020 $\pm$ 0.006 	& 	0.487 $\pm$ 0.284	&	g$_8$ = g$_7$ + g$_1$	& 	Q = 0.0671	\\		
		\hline
		p$_1$	&	8.0193	& 	0.291 $\pm$ 0.006 	& 	-1.612 $\pm$ 0.019 	&						&	Q = 0.0281	\\
		p$_2$	&	9.0588 	& 	0.591 $\pm$ 0.006 	& 	-2.940 $\pm$ 0.009 	&						&	Q = 0.0249	\\
		p$_3$	&	9.9821 	& 	0.291 $\pm$ 0.006 	& 	-3.160 $\pm$ 0.019 	&						&	Q = 0.0226	\\
		p$_4$	&	10.0324 & 	4.525 $\pm$ 0.006 	& 	1.602 $\pm$ 0.001 	&	p$_4$ = p$_5$ - g$_1$	&	Q = 0.0225	\\
		p$_5$	&	10.7273 & 	2.009 $\pm$ 0.006 	& 	-2.257 $\pm$ 0.003 	&	p$_5$ = p$_4$ + g$_1$	& 	Q = 0.0210	\\
		p$_6$	&	10.8477 & 	1.036 $\pm$ 0.006 	& 	0.065 $\pm$ 0.005 	&	p$_6$ = p$_4$ + g$_2$	&	Q = 0.0208	\\
		p$_7$	&	11.4421 & 	0.205 $\pm$ 0.006 	& 	-1.008 $\pm$ 0.027 	&						&	Q = 0.0197	\\
		p$_8$	&	16.4530 & 	0.144 $\pm$ 0.006 	& 	0.382 $\pm$ 0.034 	&						&	Q = 0.0137	\\
		\hline
		p$_{\rm mod}$	&	13.3942	&	1.444 $\pm$ 0.006	&	-1.007 $\pm$ 0.004	&	p$_{\rm mod}$ = p$_4$ + g$_8$ 		&	Q = 0.0168	\\
					&			&					&					&	p$_{\rm mod}$ = p$_5$ + g$_7$ 		&	\\
					&			&					&					&	p$_{\rm mod}$ = p$_6$ + g$_5$		&	\\
		\hline
		\end{tabular}
	\label{table: frequencies}
\end{table*}
	


	\begin{figure*}
	\centering
		\includegraphics[width=0.49\linewidth,angle=0]{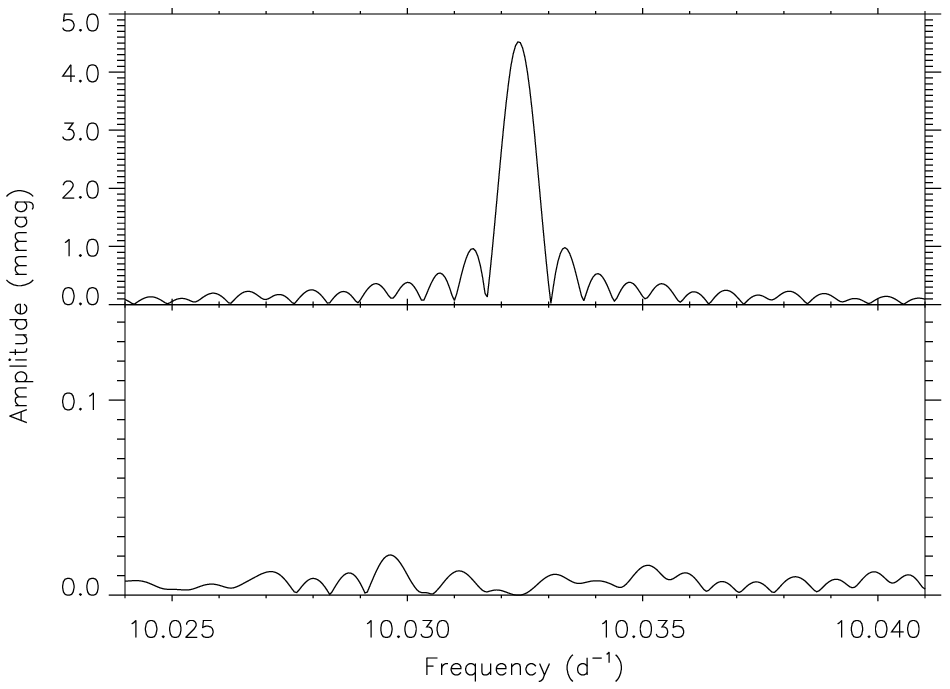}
		\includegraphics[width=0.49\linewidth,angle=0]{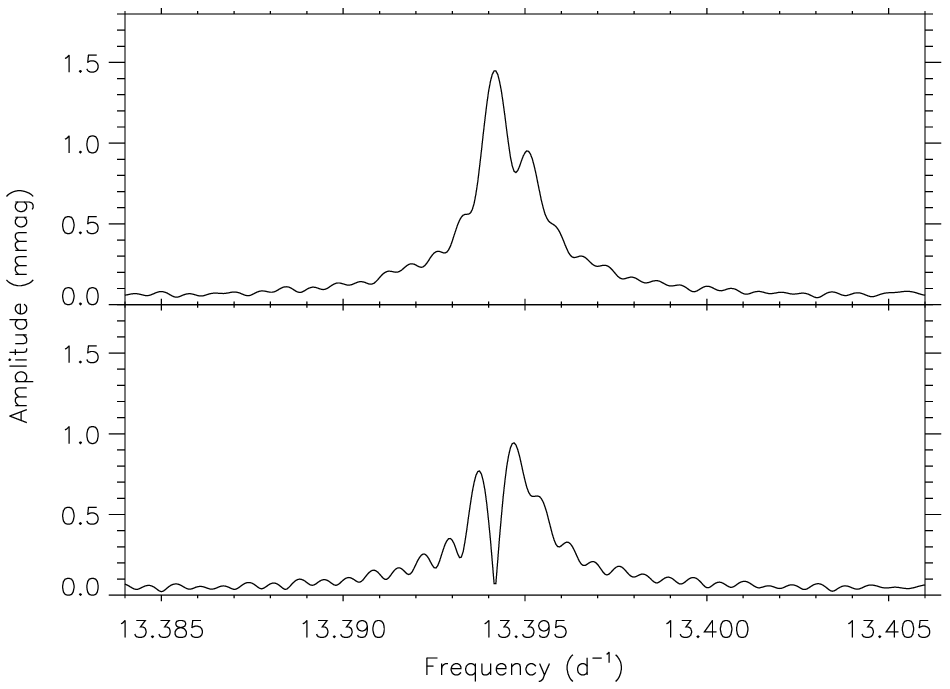}
	\caption{The left panel shows the amplitude spectrum for the strongest stable pulsation mode at $\nu$\,=\,10.0324\,d$^{-1}$ and the right panel shows the amplitude spectrum for the modulated pulsation mode at $\nu$\,=\,13.3942\,d$^{-1}$. The top and bottom parts of each panel show the amplitude spectrum before and after prewhitening respectively for each pulsation frequency. Note the change of y-axis scale for the stable mode and the fixed y-axis scale for the modulated mode.}
	\label{figure: modes}
	\end{figure*}
	

\section{A Fourier View}
\label{section: A Fourier View}

When analysing asteroseismic data, a normal approach is to use all data in a single Fourier transform, which maximises the signal-to-noise ratio and frequency resolution. In the following subsections, we explain the differences and advantages between using all data simultaneously compared to dividing the data into time bins, when performing a Fourier analysis of \textit{Kepler} data.

All the amplitude spectra produced in the analysis of KIC\,7106205 were discrete Fourier transforms and the method for calculating these numerically is described in detail in \citet{Deeming1975}. The midpoint of the entire time series, $t_0 = $\,2\,455\,688.77 (BJD), was chosen as the zero point in time when calculating every amplitude spectrum, in order to estimate phase errors accurately over 1470\,d when applying nonlinear least-squares fitting in other parts of the analysis. For the time-based approach, a bin length of 50\,d was chosen, which was a compromise between temporal and frequency resolution. Note that changing the bin length to another value between 30 and 100\,d did not significantly alter the results, but only the number of data points in Fig.\,\ref{figure: plots fixed freq}. The midpoint of each 50\,d time bin was used as the time value label on the x-axis in all plots. The step size in frequency in the amplitude spectrum was defined as $1/(20\,T)$, where $T$ is the length of each dataset in days, so that every peak was fully resolved. All of the above steps ensured that the resultant values of frequency, amplitude and phase were reliable and consistent throughout the analysis, but also that they were of sufficient accuracy to have been used as input into a least-squares fit.

\subsection{All Data Simultaneously}
\label{subsection: All Quarters}

	The amplitude spectrum for all LC data for KIC\,7106205 is shown in the bottom panel of Fig.\,\ref{figure: all quarters}. As discussed in section \ref{subsection: Hybrid Stars}, there are two frequency regimes: g\,modes have frequencies $\nu<4\,$d$^{-1}$ and p\,modes have frequencies $\nu > 5\,$d$^{-1}$. The amplitude spectrum of KIC\,7106205 contains many p\,modes that are clearly visible because of their high amplitudes compared to the noise level. Due to these strong p\,mode amplitudes, the g\,modes were drowned in the noise and associated window patterns of the p\,modes. In order to see the g\,modes easily, frequencies outside of the g\,mode regime were pre-whitened down to 3\,$\umu$mag. The strongest 8 g\,modes and 9 p\,modes were used throughout the analysis of KIC\,7106205 and are given in Table\,\ref{table: frequencies}. Since this was not a peak bagging exercise, through out this paper we refer to \textit{all peaks} as the 17 peaks given in Table\,\ref{table: frequencies}, as these have strong amplitudes for their respective frequency regime, i.e. A $\geq$ 0.1\,mmag for p\,modes and A $\geq$ 0.01\,mmag for g\,modes.
	
	The most prominent pulsation mode has a frequency of 10.0324\,d\,$^{-1}$ with an amplitude of 4.525\,$\pm$\,0.022\,mmag. In order to establish whether this mode was stable in frequency and amplitude, the peak was prewhitened with a sinusoidal function, given as equation \ref{equation: ls fitting}, where $y$ is the resultant prewhitened function, $A_i$ is the amplitude, $\nu_{i}$ is the optimised frequency obtained from the nonlinear least-squares fit, $\phi_i$ is the phase in radians, and $t$ is the time normalised to the centre of the dataset, specifically $t_0 = $\,2\,455\,688.77 (BJD). The index $i$ refers to a particular mode.
	
	\begin{equation}
		y= A_{i} \cos (2\pi\nu_{i}(t-t_0) + \phi_{i})
	\label{equation: ls fitting}
	\end{equation} 

	The amplitude spectrum of the residuals of this fit was calculated and used to examine what periodicity remained within the data after a frequency had been removed from the associated time series. This process was performed for all peaks and it was found that all modes were stable down to at most 10\,$\umu$mag, except for $\nu$\,=\,13.3942\,d$^{-1}$. A resultant prewhitened light curve is given in the top right panel of Fig.\,\ref{figure: all quarters}, in which the significant amplitude modulation of the remaining mode can be seen.
	
	In the prewhitened amplitude spectrum, the pulsation mode at $\nu$\,=\,13.3942\,d$^{-1}$ is not a resolved single peak, which is shown on the right side of Fig.\,\ref{figure: modes}. For comparison, the strongest stable frequency at $\nu$\,=\,10.0324\,d$^{-1}$ is shown on the left side of Fig.\,\ref{figure: modes}. The top and bottom panels are before and after prewhitening each peak respectively. The remaining mode amplitude after prewhitening the peak at $\nu$\,=\,13.3942\,d$^{-1}$ was approximately 1\,mmag, which was the result of the peak being unresolved over the length of the 1470\,d dataset. A single frequency value output from a least-squares fit of all the data was unable to remove the various frequencies and corresponding amplitudes produced by this mode over 1470\,d. Hence, mode amplitude remains after prewhitening -- a clear result of amplitude modulation. Since the p\,mode changed in frequency and amplitude over the length of the dataset, when all data were used in the amplitude spectrum, an unresolved peak was seen. It is important to note that this is the only pulsation mode that showed this behaviour and all others were stable in frequency and amplitude, remarkably stable to parts-per-million. {\it This establishes that $\delta$\,Sct stars can have modes stable to this high precision over 4\,yr.}
	
	\subsection{Dividing the Data into 50\,d Segments}
	\label{subsection: Quarter by Quarter}
	 
\begin{figure*}
	\centering	
		\includegraphics[width=0.49\linewidth,angle=0]{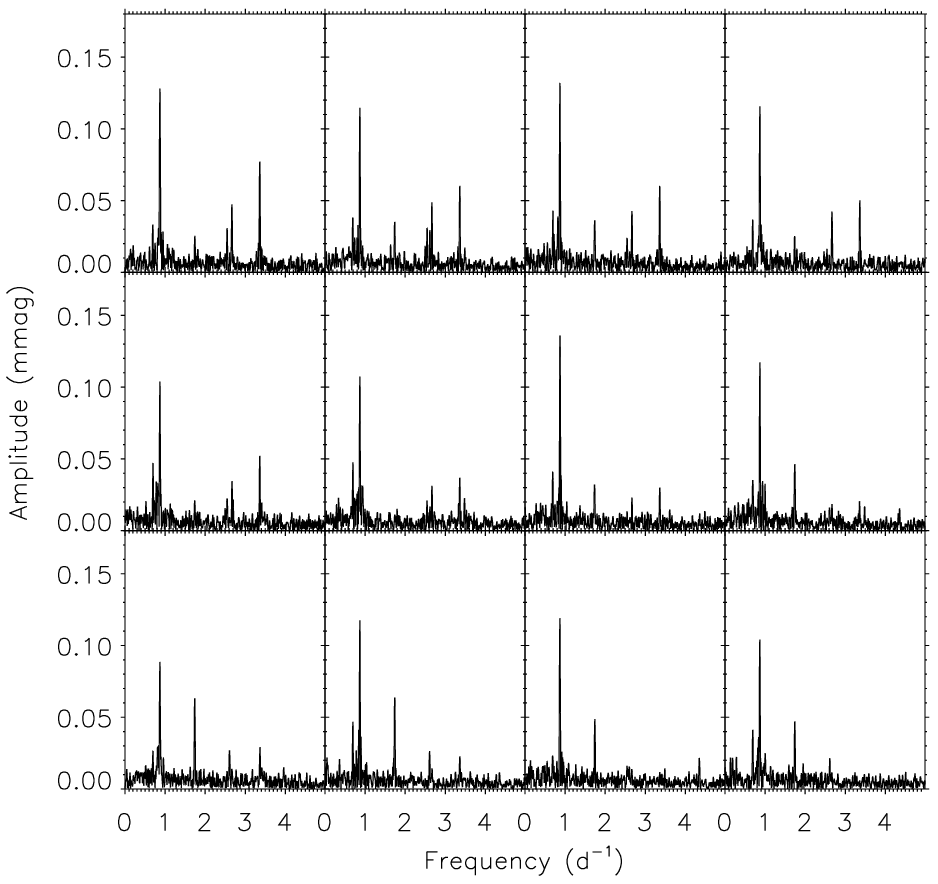}	
		\includegraphics[width=0.49\linewidth,angle=0]{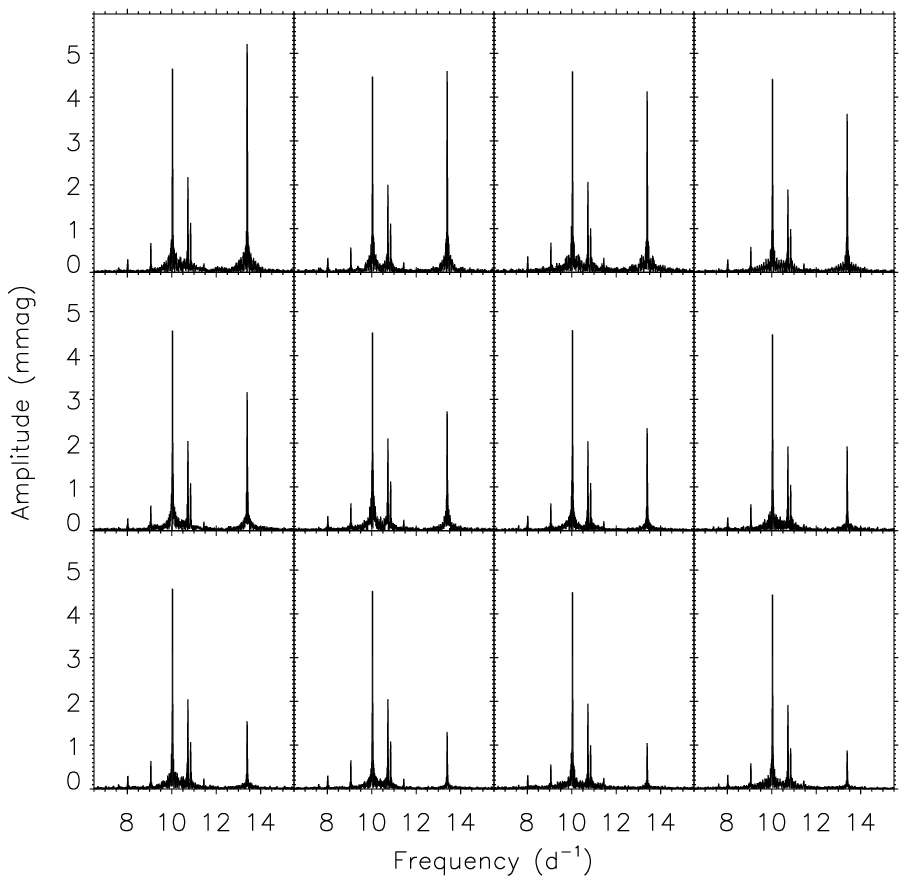}			
	\caption{Going from left to right, row by row, the left and right panels show the amplitude spectra for consecutive bins of 50\,d of \textit{Kepler} data for KIC\,7106205 for the g\,mode and p\,mode regimes respectively. The y-axis range is kept fixed in both panels, in order to show the slight modulation in g\,mode amplitudes and the significant reduction in amplitude of the p\,mode at $\nu$\,=\,13.3942\,d$^{-1}$ in each successive amplitude spectrum. All other p\,modes remain stable in frequency and amplitude.}
	\label{figure: FT year by year}
	\end{figure*}

	The data were divided into 30 bins, each 50\,d (except for the last one which was 20\,d) in length, which allowed amplitude and phase to be tracked for all modes against time. An amplitude spectrum was calculated for each bin, and the resultant values of amplitude and phase optimised using a least-squares fit with fixed frequency. This showed that the amplitude of the p\,mode at 13.3942\,d$^{-1}$ decreased over the 1470\,d dataset, from 5.161\,$\pm$\,0.031 to 0.528\,$\pm$\,0.055\,mmag. This is clearly demonstrated in the right panel of Fig.\,\ref{figure: FT year by year}, in which a section of the amplitude spectrum containing the modulated p\,mode and other stable modes was tracked throughout the entire dataset. The left panel of Fig.\,\ref{figure: FT year by year} shows the g\,mode regime tracked throughout the 1470\,d dataset. The g\,modes do show small variations in amplitude, but these are negligible. They also show no phase variation and thus are considered to be stable in both amplitude and phase, as any amplitude modulation is within the errors calculated from the least-squares fit.

	
\begin{figure*}
	\centering
		\includegraphics[width=0.49\linewidth,angle=0]{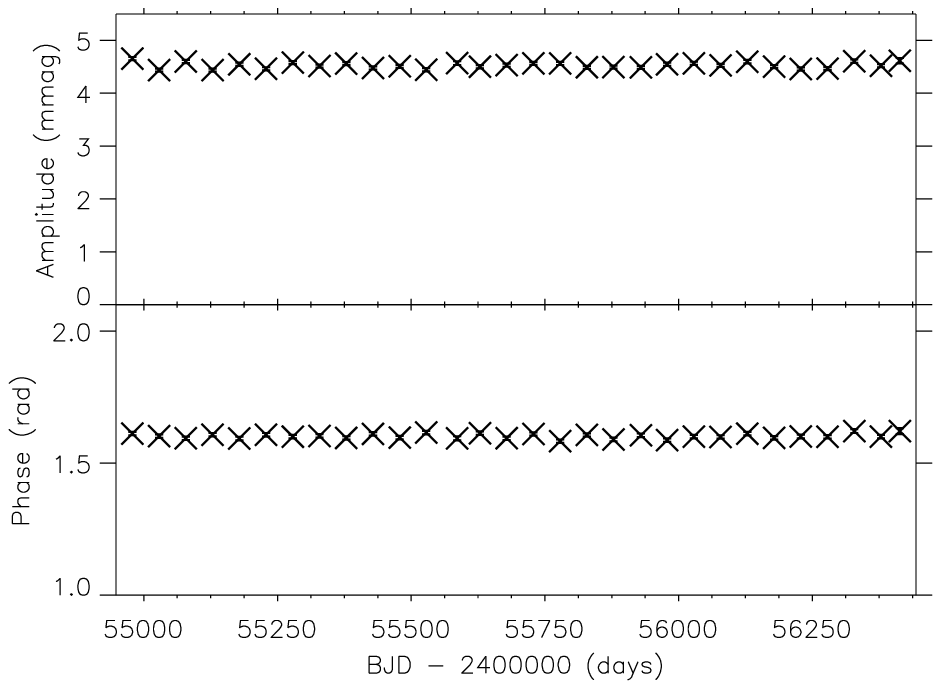}
		\includegraphics[width=0.49\linewidth,angle=0]{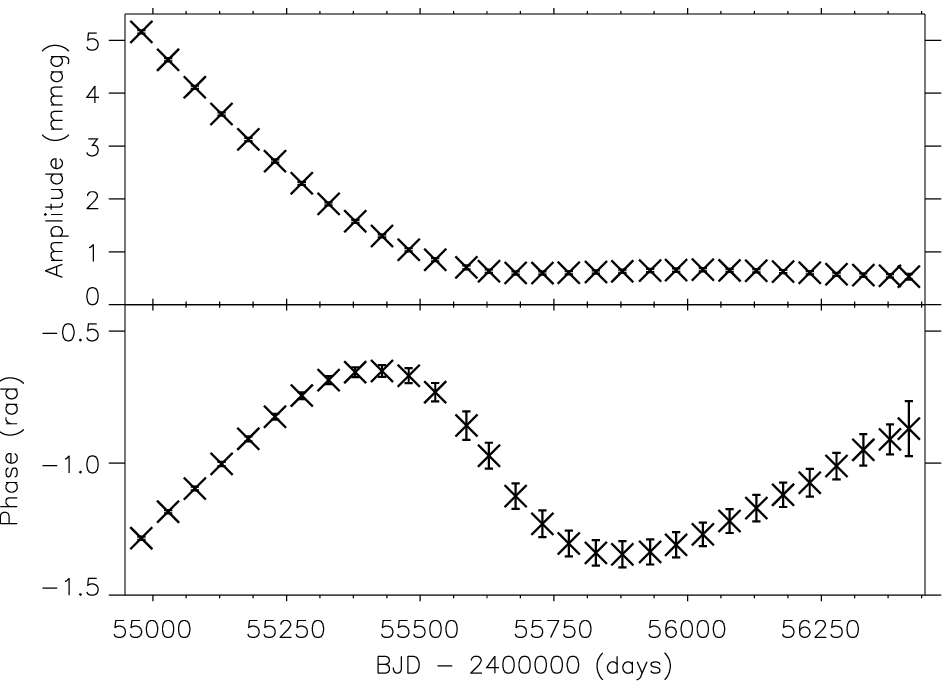}
	\caption{The left panel contains amplitude and phase plots for the strongest stable mode at a fixed frequency of 10.0324\,d$^{-1}$ over 1470\,d. The right panel contains amplitude and phase plots for the modulated mode at a fixed frequency of 13.3942\,d$^{-1}$ over 1470\,d. The x-axis is the midpoint of each 50\,d time bin in BJD -- 2\,400\,000\,d and phase values were calculated relative to the midpoint of the 1470\,d time series. 1$\sigma$ uncertainties for amplitude and phase values are also plotted, which were calculated from the least-squares fitting routine. The error bars are generally smaller than the data points. Note the y-axis scales are the same for both modes.}
	\label{figure: plots fixed freq}
	\end{figure*}

	The fixed frequency for each mode used for tracking amplitude and phase throughout the dataset was obtained from a least-squares fit using the frequency, amplitude and phase of the peak from the high-resolution amplitude spectrum for all data as input parameters. The values of amplitude and phase were then optimised using this fixed frequency for each time bin in turn. The corresponding amplitude and phase values for the mode at 13.3942\,d$^{-1}$ for each 50\,d bin of data are given in Table\,\ref{table: LS fit} and shown graphically in the right column of Fig.\,\ref{figure: plots fixed freq}. The same method for tracking amplitude and phase for all other frequencies given in Table \ref{table: frequencies} was performed. The results for the mode at 10.0324\,d$^{-1}$ are included in the left column of Fig.\,\ref{figure: plots fixed freq} for comparison.  1$\sigma$ uncertainties for values of amplitude and phase are given in Tables\,\ref{table: frequencies} and \ref{table: LS fit}, and shown in Fig.\,\ref{figure: plots fixed freq}, which were calculated from the least-squares fitting routine.

\begin{table}
\centering
\caption[]{Amplitude and phase values optimised using a least-squares fit analysis for a fixed frequency of 13.3942\,d$^{-1}$ for each 50\,d time bin, where N is the number of data points in each time bin. The phases are all with respect to the time zero point $t_0 =$\,2\,455\,688.77 (BJD). }
\small
	\begin{tabular}{c c c c}
	\hline
	\multicolumn{1}{c}{Midpoint of Bin} & \multicolumn{1}{c}{N} & \multicolumn{1}{c}{Amplitude} & \multicolumn{1}{c}{Phase} \\
				
	\multicolumn{1}{c}{BJD - 2400000 (d)} &\multicolumn{1}{c}{} & \multicolumn{1}{c}{(mmag)} & \multicolumn{1}{c}{(rad)} \\
	\hline	
				
	54978.53	&	2121		&	5.161 $\pm$ 0.031		&	-1.284 $\pm$ 0.006 \\
	55028.53	&	2258		&	4.634 $\pm$ 0.030		&	-1.184 $\pm$ 0.006 \\
	55078.53	&	2208		&	4.113 $\pm$ 0.030		&	-1.096 $\pm$ 0.007 \\
	55128.53	&	2339		&	3.607 $\pm$ 0.029		&	-1.003 $\pm$ 0.008 \\
	55178.53	&	2179		&	3.126 $\pm$ 0.030		&	-0.908 $\pm$ 0.010 \\
	55228.53	&	2178		&	2.716 $\pm$ 0.030		&	-0.824 $\pm$ 0.011 \\
	55278.53	&	2283		&	2.295 $\pm$ 0.030		&	-0.744 $\pm$ 0.013 \\
	55328.54	&	2287		&	1.907 $\pm$ 0.030		&	-0.686 $\pm$ 0.016	\\
	55378.54	&	2315		&	1.576 $\pm$ 0.029		&	-0.656 $\pm$ 0.019	\\
	55428.54	&	2379		&	1.300 $\pm$ 0.029		&	-0.651 $\pm$ 0.022	\\
	55478.54	&	2312		&	1.037 $\pm$ 0.029		&	-0.669 $\pm$ 0.028	\\
	55528.05	&	2307		&	0.850 $\pm$ 0.029		&	-0.731 $\pm$ 0.035	\\
	55586.43	&	1391		&	0.705 $\pm$ 0.038		&	-0.858 $\pm$ 0.054	\\
	55628.54	&	2078		&	0.633 $\pm$ 0.031		&	-0.972 $\pm$ 0.049	\\
	55678.54	&	2390		&	0.599 $\pm$ 0.029		&	-1.125 $\pm$ 0.048	\\
	55728.55	&	2200		&	0.598 $\pm$ 0.030		&	-1.230 $\pm$ 0.050	\\
	55777.89	&	2310		&	0.603 $\pm$ 0.029		&	-1.304 $\pm$ 0.049	\\
	55828.66	&	2272		&	0.618 $\pm$ 0.030		&	-1.340 $\pm$ 0.048	\\
	55878.10	&	2031		&	0.632 $\pm$ 0.031		&	-1.345 $\pm$ 0.050	\\
	55929.93	&	2116		&	0.645 $\pm$ 0.031		&	-1.336 $\pm$ 0.048	\\
	55978.55	&	2077		&	0.653 $\pm$ 0.031		&	-1.309 $\pm$ 0.048	\\
	56028.54	&	2332		&	0.661 $\pm$ 0.029		&	-1.270 $\pm$ 0.044	\\
	56078.52	&	2335		&	0.647 $\pm$ 0.029		&	-1.219 $\pm$ 0.045	\\
	56128.53	&	1945		&	0.640 $\pm$ 0.032		&	-1.170 $\pm$ 0.050	\\
	56178.54	&	2379		&	0.626 $\pm$ 0.029		&	-1.120 $\pm$ 0.046	\\
	56228.54	&	1985		&	0.603 $\pm$ 0.032		&	-1.074 $\pm$ 0.053	\\
	56278.53	&	2387		&	0.575 $\pm$ 0.029		&	-1.011 $\pm$ 0.050	\\
	56328.53	&	1808		&	0.558 $\pm$ 0.033		&	-0.950 $\pm$ 0.059	\\
	56378.54	&	2062		&	0.545 $\pm$ 0.031		&	-0.910 $\pm$ 0.057	\\
	56413.77	&	663		&	0.528 $\pm$ 0.055		&	-0.869 $\pm$ 0.104	\\

	\hline
	\end{tabular}
\label{table: LS fit}
\end{table}

It is clear that the amplitude of the 13.3942\,d$^{-1}$ mode dropped significantly and the phase changed over the length of the data set, while all other mode frequencies were stable in both amplitude and phase. We now discuss the implications of these observations.

	
	\section{Discussion}
	\label{section: Discussion}
	
	The hypothesis proposed in section \ref{section: Amplitude Modulation} concerning amplitude modulation has been successfully tested: \textit{Is the visible pulsation energy conserved in KIC\,7106205?} 
	
	Small changes in amplitude were observed in the g\,modes, but these changes were negligible as the errors on amplitudes were relatively much larger on smaller values. Moreover, the changes in g\,mode amplitudes were not sufficient to explain the significant decrease in amplitude of the mode at $\nu$\,=\,13.3942\,d$^{-1}$. Since the modulated mode was decreasing in amplitude, one would expect the g\,mode amplitudes to increase, if energy were to be conserved between the visible pulsation modes. Since, the negligible variation in the g\,mode amplitudes also show a slight decrease in amplitude over the dataset, the energy from the mode at $\nu$\,=\,13.3942\,d$^{-1}$ is not being transferred to the g\,modes.
	
	Table\,\ref{table: frequencies} lists the frequencies, amplitudes and phases with respective errors of the g and p\,modes used in this analysis. It also labels the frequencies most likely to be combination frequencies. The pulsation constants were calculated using equation \ref{equation: pulsation constant},
	
	\begin{equation}
		\log\rm{Q} = \log\rm{P} + \frac{1}{2} \log g + \frac{1}{10} M_{\rm Bol} + \log\,T_{\rm eff} - 6.454
	\label{equation: pulsation constant}
	\end{equation} 		
	
\noindent	where P is the pulsation period in days, $M_{\rm Bol}$ is the bolometric magnitude (a value of $M_{\rm Bol} = 1.2$ was calculated using T$_{\rm eff}$ and $\log g$ in comparison with the Pleiades main sequence) and other parameters have their usual meanings with values taken from \citet{Huber2014}. The modulated p\,mode at $\nu$\,=\,13.3942\,d$^{-1}$ is subject to combination effects, but since the g\,mode amplitudes are so small and their variation is negligible, it is unlikely that these combination effects are the cause of the significant amplitude modulation. The strongest pulsation frequency at $\nu$\,=\,10.0324\,d$^{-1}$ has a pulsation constant of Q = 0.0225\,d, which suggests that is a first or second radial overtone mode (see \citealt{Stellingwerf1979}). The modulated mode at $\nu$\,=\,13.3942\,d$^{-1}$ has a pulsation constant of Q = 0.0168\,d which suggests it is a third or fourth radial overtone mode. Since none of the p\,modes have Q-values greater than the 0.033\,d expected for the radial fundamental mode, there is no evidence of any mixed modes.
	
	From the amplitude spectrum of all data, we found that the amplitudes of the 8 highest p and 8 highest g\,modes were stable to at most 10\,$\umu$mag, and a single modulated mode at 13.3942\,d$^{-1}$ was unstable. Amplitude and phase values, optimised by least squares fitting, for a fixed frequency were also tracked across 1470\,d and found to be stable for all modes except $\nu$\,=\,13.3942\,d$^{-1}$. The most important point to note is that KIC\,7106205 is a star that contains pulsation modes, stable in frequency and amplitude to remarkable levels of precision, but also contains a single pulsation mode that is not. This behaviour is astrophysical and is not an instrumental effect. If this were an instrumental effect, one would expect all modes to decrease (or increase) in amplitude by the same amount over the length of the dataset. Since a significant decrease in amplitude of the pulsation mode at $\nu$\,=\,13.3942\,d$^{-1}$ was observed and all other modes were stable in frequency and amplitude, it is concluded that the visible pulsation energy was not conserved. In the following subsections, the possible causes of the amplitude modulation and quasi-sinusoidal phase variation are discussed.
	
	\subsection{Causes of Decreasing Mode Amplitude}
	
	Analysis of KIC\,7106205 showed a single mode that decreased in amplitude by an order of magnitude, on a timescale less than 1470\,d, with all other modes remaining stable in amplitude and phase. This star is an example of a hybrid star within the \textit{Kepler} dataset with p and g\,modes, but also both stable and unstable pulsation modes, among which visible energy is not conserved. The energy from this modulated mode has gone somewhere other than into the visible pulsation modes, as all other modes were observed to be stable in frequency and amplitude. The following are possible causes of the amplitude modulation observed in KIC\,7106205.
	
	\begin{enumerate}
	
	\item \textit{Energy is lost to the convection or ionisation zone:} The energy may have been lost to the convection zone or He\,\textsc{ii} ionisation zone, both of which are driving regions in $\delta$\,Sct and $\gamma$\,Dor stars. This suggests that driving and damping are unstable processes on the time scale studied for some pulsation modes, but not for others. We conjecture that this is a function of the structure of the mode cavity with respect to the driving zones, particularly the positions of the radial nodes. Modelling studies might shed some light on this conjecture, and further observational studies of the \textit{Kepler} hybrid stars will illuminate the range of mode stability characteristics observationally. Similar behaviour was reported by \citet{Breger2000a} from ground-based study of the $\delta$\,Sct star 4\,CVn. KIC\,7106205 demonstrates how much more clearly we can examine mode stability with the nearly continuous, high precision \textit{Kepler} 4\,yr data.
	
	\item \textit{Mode Beating:} The change in amplitude of the modulated p\,mode at $\nu$\,=\,13.3942\,d$^{-1}$ is unlikely to be the result of unresolved frequencies beating. The amplitude spectrum of the full 1470\,d data set shows a broad peak not consistent with only two unresolved modes, hence a hypothesis of beating of unresolved mode frequencies would require more than two. That is further supported by the apparently non-periodic behaviour of the amplitude variability, coupled with the quasi-sinusoidal phase variability. This case is not similar to that of 4\,CVn, where \citet{Breger2000b} did suggest beating of unresolved mode frequencies. Finally, for mode frequencies to be unresolved in the 1470\,d full {\it Kepler} data set, they have to be closer in frequency than 8\,nHz. It is an unresolved question whether two mode frequencies can be this close in a pulsating star and still maintain their independent characters. Modelling studies are needed to address this question.
		
	\item \textit{Stellar Evolution:} Pulsations are the result of the balance between the driving and damping mechanisms in a star. Could the modulation of the $\nu$\,=\,13.3942\,d$^{-1}$ mode be the result of evolutionary changes in the structure of KIC\,7106205, which lies close to the red edge of the $\delta$\,Sct instability strip \citep{Uytterhoeven2011}? The answer to that question is not known, but for it to be true, the pulsation cavities would need to be finely tuned for all other modes to show the high level of stability observed.

	\item \textit{Mode-Coupling:} As mentioned in section\,\ref{section: Amplitude Modulation}, theory of $\delta$\,Sct star pulsations predicts mode-coupling that can affect pulsation frequencies. \citet{Dziembowski1982} derived a parametric resonance condition of $\nu_1 \approx \nu_2 + \nu_3$, in which a low radial order, linearly driven mode, $\nu_1$, is able to decay into two other modes, $\nu_2$ and $\nu_3$. The most likely outcome is two g\,modes with low amplitudes. If the g\,modes are not observable, then this hypothesis cannot be verified. Small and negligible changes in amplitude were seen in the low-amplitude g\,modes present in KIC\,7106205. However, the only combination frequencies that included the mode at $\nu$\,=\,13.3942\,d$^{-1}$ do not exclusively involve two g\,modes that satisfy $\nu_1 \approx \nu_2 + \nu_3$. Moreover, the small amplitudes of the observed g\,modes cannot explain the amplitude modulation seen in the mode at $\nu$\,=\,13.3942\,d$^{-1}$ according to the parametric resonance theory of \citet{Dziembowski1982}, as the amplitudes of the coupled modes need to be comparable. No growth of g\,mode amplitudes was observed. \citet{Breger2014} do observe parent-daughter mode-coupling in {\it Kepler} data of the unusual $\delta$\,Sct star KIC\,8054146, but we see no similar behaviour in KIC\,7106205. We consider it unlikely that mode-coupling to g\,modes that are not visible in the amplitude spectrum is an explanation for the changes in the 13.3942\,d$^{-1}$ mode. However, we cannot rule our coupling to high degree (high $\ell$) modes that are not visible in broadband photometry. A hypothesis of mode-coupling to unobservable high degree modes is not testable with {\it Kepler} data. This does not contradict our conclusion that we have tested the hypothesis of energy conservation in {\it observable} modes.
	
	\end{enumerate}
	
	\subsection{Quasi-sinusoidal Phase Variations}
	
	A remarkable outcome of this analysis was the sinusoidal variation in phase of the p\,mode at a fixed frequency of $\nu$\,=\,13.3942\,d$^{-1}$. This cannot be the result of frequency modulation in a binary system \citep{Shibahashi2012}, since that would affect all pulsation frequencies and would not generate amplitude modulation. The changes in the 13.3942\,d$^{-1}$ mode are internal to the star. A magnetic cycle with changes confined to the thin surface convection zone might conceivably affect only the highest radial overtone mode. However, the existence of lower amplitude modes with even higher frequencies, such as the one at $\nu$\,=\,16.4530\,d$^{-1}$ listed in Table\,\ref{table: frequencies}, argues against this.
	
	We know that the pulsation frequencies in the Sun are variable over the solar cycle \citep{Chaplin2000, Chaplin2007}, but whether such an effect is possible in $\delta$\,Sct stars is unknown. In the case of KIC\,7106205 the lack of any indication of periodicity in the amplitude modulation of the 13.3942\,d$^{-1}$ mode does not support a hypothesis of cyclic behaviour, and we note that for the phase variation there is only one cycle that may, or may not continue to vary quasi-sinusoidally. The possibly periodic nature of this variation suggests that the mode may become re-excited in the future and start to increase in amplitude once again, but with only a single cycle observed, this is very speculative. This was not observed in the amplitude and phase variations of 4\,CVn, as the re-excited pulsation mode in that star was preceded by a phase shift of half a cycle \citep{Breger2000b, Breger2009}. Quasi-sinusoidal phase variation was also observed in pulsation modes in KIC\,8054146, which \citet{Breger2014} interpreted as mode-coupling, since they could observe the coupled modes in that star. We cannot do that for KIC\,7106205. We therefore reach no conclusion about the quasi-sinusoidal phase variation of the 13.3924\,d$^{-1}$ mode, but simply note it.


\section{Conclusions}
\label{section: Conclusions}

	KIC\,7106205 is an ideal case study of a $\delta$\,Sct star that exhibits amplitude modulation because only a single pulsation mode is modulated over the length of the dataset, whilst all other modes remained highly stable in both frequency and amplitude. We showed that this amplitude modulation is astrophysical, not instrumental. A single p\,mode at $\nu$\,=\,13.3942\,d$^{-1}$ decreased in amplitude from 5.161\,$\pm$\,0.031 to 0.528\,$\pm$\,0.055\,mmag over 1470\,d. The missing mode power was not seen to increase the amplitude of any existing p or g\,mode, nor did it excite any new visible modes in either of the low or high frequency regime. Hence, we conclude that the visible pulsation energy is not conserved in KIC\,7106205.
	
	There are two possible explanations for the lost mode energy: mode-coupling or loss of mode energy to the convection zone and/or the pulsational p-mode driving He\,\textsc{ii} ionisation zone. We suggested that mode-coupling to g\,modes is not a likely explanation, as even though g\,modes are visible in hybrid stars such as KIC\,7106205, none of the g\,modes observed meet the criteria specified in \citet{Dziembowski1982}. We cannot rule out coupling to high degree modes that are not visible in broadband photometry. \citet{Breger2000b} speculated that power can be transferred between pulsation modes by a mode-coupling mechanism but was unable to completely explain the amplitude modulation observed in 4\,CVn. Similarly, \citet{Breger2014} concluded that a mode-coupling mechanism was operating in KIC\,8054146, in which amplitude and phase variation of many different modes was observed. A similar mode-coupling mechanism could be at work within KIC\,7106205 and such effects are important if one is to fully understand the mechanisms that excite pulsation modes and the physics that governs hybrid stars within the instability strip. Alternatively, if mode-coupling is not the explanation, then the lost energy must have been transferred somewhere in the star; whether this is caused by a decrease in driving in the He\,\textsc{ii} ionisation zone, or an increase in damping elsewhere is not known.

	Stellar evolution theory and convection models predict that real-time evolution of a star can be observed with a dataset as short as a few years. \cite{Murphy2012b} analysed a $\rho$\,Puppis star, in which two  pulsation modes grew in amplitude over the time span of the dataset. They concluded that the growth in mode amplitude might be the result of the pulsation driving zone being driven deeper within the He\,\textsc{ii} ionisation zone, which is predicted by stellar evolution models for this star. Therefore, the star was evolving in real time over the 2\,yr of observations. Since a decline in mode amplitude was observed in KIC\,7106205, it is unlikely that this is the same effect. It does mean that stellar evolution can operate on short timescales, on the order of years, but does not provide support for the amplitude modulation being caused by the outer layers of KIC\,7106205 altering the pulsation cavity for the 13.3942\,d$^{-1}$ mode. The mode at 13.3942\,d$^{-1}$ is not the highest overtone observed in KIC\,7106205, it is unlikely that stellar evolution is the solution to the amplitude modulation seen, as a higher frequency mode at $\nu$\,=\,16.4530\,d$^{-1}$ was observed to be stable.

	 Both mode-coupling to high-degree modes and energy being lost to a damping region could explain the amplitude modulation seen in KIC\,7106205. The analysis of the star does not distinguish which is more likely, only that the energy lost from the modulated mode was not transferred to any of the existing observed pulsation modes. In reality, both solutions could be operating in congruence, as both are feasible solutions to the observed amplitude modulation, which further complicates the issue of testing the hypothesis of energy conservation between pulsation modes.
	 
	 These conclusions are speculative, but this is a step toward understanding the phenomenon of amplitude modulation within $\delta$\,Sct and $\gamma$\,Dor stars using the \textit{Kepler} dataset. What is certain is that the hypothesis concerning energy conservation between visible pulsation modes has been successfully tested and proved not to be the case in KIC\,7106205. This problem of mode stability is widely observed in pulsating stars. For stochastically driven, solar-like and red giant pulsators there is no problem understanding changes in mode amplitude and phase, given the stochastic nature of the driving. For heat engine pulsators, the situation is unclear.


\section*{acknowledgements}
DMB wishes to thank the STFC for the financial support of his PhD and the \textit{Kepler} team for providing such excellent data.


\bibliography{KIC7106205_bib.bib}

\begin{thebibliography}{30}
\expandafter\ifx\csname natexlab\endcsname\relax\def\natexlab#1{#1}\fi

\bibitem[{{Borucki} {et~al}\mbox{.}(2010){Borucki}, {Koch}, {Basri}, {Batalha},
  {Brown}, {Caldwell}, {Caldwell}, {Christensen-Dalsgaard}, {Cochran},
  {DeVore}, {Dunham}, {Dupree}, {Gautier}, {Geary}, {Gilliland}, {Gould},
  {Howell}, {Jenkins}, {Kondo}, {Latham}, {Marcy}, {Meibom}, {Kjeldsen},
  {Lissauer}, {Monet}, {Morrison}, {Sasselov}, {Tarter}, {Boss}, {Brownlee},
  {Owen}, {Buzasi}, {Charbonneau}, {Doyle}, {Fortney}, {Ford}, {Holman},
  {Seager}, {Steffen}, {Welsh}, {Rowe}, {Anderson}, {Buchhave}, {Ciardi},
  {Walkowicz}, {Sherry}, {Horch}, {Isaacson}, {Everett}, {Fischer}, {Torres},
  {Johnson}, {Endl}, {MacQueen}, {Bryson}, {Dotson}, {Haas}, {Kolodziejczak},
  {Van Cleve}, {Chandrasekaran}, {Twicken}, {Quintana}, {Clarke}, {Allen},
  {Li}, {Wu}, {Tenenbaum}, {Verner}, {Bruhweiler}, {Barnes}, \&
  {Prsa}}]{Borucki2010}
{Borucki} W.~J. {et~al.}, 2010, Science, 327, 977

\bibitem[{{Breger}(2000{\natexlab{a}})}]{Breger2000a}
{Breger} M., 2000{\natexlab{a}}, in Astronomical Society of the Pacific
  Conference Series, Vol. 210, Delta Scuti and Related Stars, {Breger} M.,
  {Montgomery} M., eds., p.~3

\bibitem[{{Breger}(2000{\natexlab{b}})}]{Breger2000b}
---, 2000{\natexlab{b}}, \mnras, 313, 129

\bibitem[{{Breger}(2009)}]{Breger2009}
---, 2009, in American Institute of Physics Conference Series, Vol. 1170,
  American Institute of Physics Conference Series, {Guzik} J.~A., {Bradley}
  P.~A., eds., pp. 410--414

\bibitem[{{Breger} \& {Montgomery}(2014)}]{Breger2014}
{Breger} M., {Montgomery} M.~H., 2014, \apj, 783, 89

\bibitem[{{Brown} {et~al}\mbox{.}(2011){Brown}, {Latham}, {Everett}, \&
  {Esquerdo}}]{Brown2011}
{Brown} T.~M., {Latham} D.~W., {Everett} M.~E., {Esquerdo} G.~A., 2011, \aj,
  142, 112

\bibitem[{{Chaplin} {et~al}\mbox{.}(2000){Chaplin}, {Elsworth}, {Isaak},
  {Miller}, \& {New}}]{Chaplin2000}
{Chaplin} W.~J., {Elsworth} Y., {Isaak} G.~R., {Miller} B.~A., {New} R., 2000,
  \mnras, 313, 32

\bibitem[{{Chaplin} {et~al}\mbox{.}(2007){Chaplin}, {Elsworth}, {Miller},
  {Verner}, \& {New}}]{Chaplin2007}
{Chaplin} W.~J., {Elsworth} Y., {Miller} B.~A., {Verner} G.~A., {New} R., 2007,
  \apj, 659, 1749

\bibitem[{{Chevalier}(1971)}]{Chevalier1971a}
{Chevalier} C., 1971, \aap, 14, 24

\bibitem[{{Deeming}(1975)}]{Deeming1975}
{Deeming} T.~J., 1975, \apss, 36, 137

\bibitem[{{Dupret} {et~al}\mbox{.}(2004){Dupret}, {Grigahc{\`e}ne}, {Garrido},
  {Gabriel}, \& {Scuflaire}}]{Dupret2004}
{Dupret} M.-A., {Grigahc{\`e}ne} A., {Garrido} R., {Gabriel} M., {Scuflaire}
  R., 2004, \aap, 414, L17

\bibitem[{{Dupret} {et~al}\mbox{.}(2005){Dupret}, {Grigahc{\`e}ne}, {Garrido},
  {Gabriel}, \& {Scuflaire}}]{Dupret2005}
---, 2005, \aap, 435, 927

\bibitem[{{Dziembowski}(1982)}]{Dziembowski1982}
{Dziembowski} W., 1982, Acta Astronomica, 32, 147

\bibitem[{{Gilliland} {et~al}\mbox{.}(2010){Gilliland}, {Brown},
  {Christensen-Dalsgaard}, {Kjeldsen}, {Aerts}, {Appourchaux}, {Basu},
  {Bedding}, {Chaplin}, {Cunha}, {De Cat}, {De Ridder}, {Guzik}, {Handler},
  {Kawaler}, {Kiss}, {Kolenberg}, {Kurtz}, {Metcalfe}, {Monteiro}, {Szab{\'o}},
  {Arentoft}, {Balona}, {Debosscher}, {Elsworth}, {Quirion}, {Stello},
  {Su{\'a}rez}, {Borucki}, {Jenkins}, {Koch}, {Kondo}, {Latham}, {Rowe}, \&
  {Steffen}}]{Gilliland2010}
{Gilliland} R.~L. {et~al.}, 2010, \pasp, 122, 131

\bibitem[{{Grigahc{\`e}ne} {et~al}\mbox{.}(2005){Grigahc{\`e}ne}, {Dupret},
  {Gabriel}, {Garrido}, \& {Scuflaire}}]{Griga2005}
{Grigahc{\`e}ne} A., {Dupret} M.-A., {Gabriel} M., {Garrido} R., {Scuflaire}
  R., 2005, \aap, 434, 1055

\bibitem[{{Grigahc{\`e}ne} {et~al}\mbox{.}(2010){Grigahc{\`e}ne},
  {Uytterhoeven}, {Antoci}, {Balona}, {Catanzaro}, {Daszy{\'n}ska-Daszkiewicz},
  {Guzik}, {Handler}, {Houdek}, {Kurtz}, {Marconi}, {Monteiro}, {Moya},
  {Ripepi}, {Su{\'a}rez}, {Borucki}, {Brown}, {Christensen-Dalsgaard},
  {Gilliland}, {Jenkins}, {Kjeldsen}, {Koch}, {Bernabei}, {Bradley}, {Breger},
  {Di Criscienzo}, {Dupret}, {Garc{\'{\i}}a}, {Garc{\'{\i}}a Hern{\'a}ndez},
  {Jackiewicz}, {Kaiser}, {Lehmann}, {Mart{\'{\i}}n-Ruiz}, {Mathias},
  {Molenda-{\.Z}akowicz}, {Nemec}, {Nuspl}, {Papar{\'o}}, {Roth}, {Szab{\'o}},
  {Suran}, \& {Ventura}}]{Griga2010}
{Grigahc{\`e}ne} A. {et~al.}, 2010, Astronomische Nachrichten, 331, 989

\bibitem[{{Guzik} {et~al}\mbox{.}(2000){Guzik}, {Kaye}, {Bradley}, {Cox}, \&
  {Neuforge}}]{Guzik2000a}
{Guzik} J.~A., {Kaye} A.~B., {Bradley} P.~A., {Cox} A.~N., {Neuforge} C., 2000,
  \apjl, 542, L57

\bibitem[{{Haas} {et~al}\mbox{.}(2014){Haas}, {Barclay}, {Batalha}, {Bryson},
  {Caldwell}, {Campbell}, {Coughlin}, {Howell}, {Jenkins}, {Klaus}, {Mullally},
  {Sanderfer}, {Sobeck}, {Still}, {Troeltzsch}, \& {Twicken}}]{Haas2014}
{Haas} M.~R. {et~al.}, 2014, in American Astronomical Society Meeting
  Abstracts, Vol. 223, American Astronomical Society Meeting Abstracts

\bibitem[{{Handler}(1999)}]{Handler1999}
{Handler} G., 1999, \mnras, 309, L19

\bibitem[{{Huber} {et~al}\mbox{.}(2014){Huber}, {Silva Aguirre}, {Matthews},
  {Pinsonneault}, {Gaidos}, {Garc{\'{\i}}a}, {Hekker}, {Mathur}, {Mosser},
  {Torres}, {Bastien}, {Basu}, {Bedding}, {Chaplin}, {Demory}, {Fleming},
  {Guo}, {Mann}, {Rowe}, {Serenelli}, {Smith}, \& {Stello}}]{Huber2014}
{Huber} D. {et~al.}, 2014, \apjs, 211, 2

\bibitem[{{Jenkins} {et~al}\mbox{.}(2010){Jenkins}, {Caldwell},
  {Chandrasekaran}, {Twicken}, {Bryson}, {Quintana}, {Clarke}, {Li}, {Allen},
  {Tenenbaum}, {Wu}, {Klaus}, {Van Cleve}, {Dotson}, {Haas}, {Gilliland},
  {Koch}, \& {Borucki}}]{Jenkins2010b}
{Jenkins} J.~M. {et~al.}, 2010, \apjl, 713, L120

\bibitem[{{Kaye} {et~al}\mbox{.}(2000){Kaye}, {Handler}, {Krisciunas},
  {Poretti}, \& {Zerbi}}]{Kaye2000}
{Kaye} A.~B., {Handler} G., {Krisciunas} K., {Poretti} E., {Zerbi} F.~M., 2000,
  in Astronomical Society of the Pacific Conference Series, Vol. 203, IAU
  Colloq. 176: The Impact of Large-Scale Surveys on Pulsating Star Research,
  {Szabados} L., {Kurtz} D., eds., pp. 426--429

\bibitem[{{Koch} {et~al}\mbox{.}(2010){Koch}, {Borucki}, {Basri}, {Batalha},
  {Brown}, {Caldwell}, {Christensen-Dalsgaard}, {Cochran}, {DeVore}, {Dunham},
  {Gautier}, {Geary}, {Gilliland}, {Gould}, {Jenkins}, {Kondo}, {Latham},
  {Lissauer}, {Marcy}, {Monet}, {Sasselov}, {Boss}, {Brownlee}, {Caldwell},
  {Dupree}, {Howell}, {Kjeldsen}, {Meibom}, {Morrison}, {Owen}, {Reitsema},
  {Tarter}, {Bryson}, {Dotson}, {Gazis}, {Haas}, {Kolodziejczak}, {Rowe}, {Van
  Cleve}, {Allen}, {Chandrasekaran}, {Clarke}, {Li}, {Quintana}, {Tenenbaum},
  {Twicken}, \& {Wu}}]{Koch2010}
{Koch} D.~G. {et~al.}, 2010, \apjl, 713, L79

\bibitem[{{Murphy}(2012)}]{Murphy2012a}
{Murphy} S.~J., 2012, \mnras, 422, 665

\bibitem[{{Murphy} {et~al}\mbox{.}(2012){Murphy}, {Grigahc{\`e}ne},
  {Niemczura}, {Kurtz}, \& {Uytterhoeven}}]{Murphy2012b}
{Murphy} S.~J., {Grigahc{\`e}ne} A., {Niemczura} E., {Kurtz} D.~W.,
  {Uytterhoeven} K., 2012, \mnras, 427, 1418

\bibitem[{{Murphy}, {Shibahashi} \& {Kurtz}(2013){Murphy}, {Shibahashi}, \&
  {Kurtz}}]{Murphy2013}
{Murphy} S.~J., {Shibahashi} H., {Kurtz} D.~W., 2013, \mnras, 430, 2986

\bibitem[{{Rodr{\'{\i}}guez} \& {Breger}(2001)}]{Rod2001}
{Rodr{\'{\i}}guez} E., {Breger} M., 2001, \aap, 366, 178

\bibitem[{{Shibahashi} \& {Kurtz}(2012)}]{Shibahashi2012}
{Shibahashi} H., {Kurtz} D.~W., 2012, \mnras, 422, 738

\bibitem[{{Stellingwerf}(1979)}]{Stellingwerf1979}
{Stellingwerf} R.~F., 1979, \apj, 227, 935

\bibitem[{{Uytterhoeven} {et~al}\mbox{.}(2011){Uytterhoeven}, {Moya},
  {Grigahc{\`e}ne}, {Guzik}, {Guti{\'e}rrez-Soto}, {Smalley}, {Handler},
  {Balona}, {Niemczura}, {Fox Machado}, {Benatti}, {Chapellier}, {Tkachenko},
  {Szab{\'o}}, {Su{\'a}rez}, {Ripepi}, {Pascual}, {Mathias},
  {Mart{\'{\i}}n-Ru{\'{\i}}z}, {Lehmann}, {Jackiewicz}, {Hekker},
  {Gruberbauer}, {Garc{\'{\i}}a}, {Dumusque}, {D{\'{\i}}az-Fraile}, {Bradley},
  {Antoci}, {Roth}, {Leroy}, {Murphy}, {De Cat}, {Cuypers}, {Kjeldsen},
  {Christensen-Dalsgaard}, {Breger}, {Pigulski}, {Kiss}, {Still}, {Thompson},
  \& {van Cleve}}]{Uytterhoeven2011}
{Uytterhoeven} K. {et~al.}, 2011, \aap, 534, A125

\end{thebibliography}


\end{document}